\documentclass[aps,prl,twocolumn,groupedaddress]{revtex4}

\usepackage{amsmath}
\usepackage{hyperref}
\usepackage{slashed}
\usepackage{graphics}
\usepackage{bm,bbm}

\begin{document}


\title{Quantum Gravity and Causal Structures:\\[1mm]  Second Quantization of Conformal Dirac Algebras}


\author{R.~Bonezzi}
\email[]{bonezzi@bo.infn.it}
\affiliation{
Dipartimento di Fisica,Universit\`a di Bologna, via Irnerio 46, I-40126 Bologna, Italy
and INFN sezione di Bologna, via Irnerio 46, I-40126 Bologna, Italy, and
Departamento de Ciencias F\'isicas,
Universidad Andres Bello, Republica 220, Santiago, Chile
}

\author{O.~Corradini}
\email[]{olindo.corradini@unach.mx}
\affiliation{Facultas de Ciencias en F\'isica y Matem\'aticas, Universidad Aut\'onoma de Chiapas,
          Ciudad Universitaria, Tuxtla Guti\'errez 29050, M\'exico
          and Dipartimento di Scienze Fisiche, Informatiche e Matematiche,\\ Universit\`a di Modena e Reggio Emilia, Via Campi 213/A,  I-41125 Modena, Italy}

\author{E.~Latini}
\email[]{emanuele.latini@math.uzh.ch}
\affiliation{Institut f{\"u}r Mathematik, Universit{\"a}t Z{\"u}rich-Irchel, Winterthurerstrasse 190, CH-8057 Z{\"u}rich,
  Switzerland, and INFN, Laboratori Nazionali di Frascati, CP 13,
  I-00044 Frascati, Italy}

\author{A.~Waldron}
\email[]{wally@math.ucdavis.edu}
\affiliation{Department of Mathematics, University of California, Davis, CA 95616, USA}


\date{\today}

\begin{abstract}
It is postulated that quantum gravity is a sum over causal structures   coupled  to matter via scale evolution.  Quantized causal structures can be described by studying simple matrix models where matrices are replaced by an algebra of quantum mechanical observables. In particular, previous studies constructed quantum gravity models by quantizing the moduli  of Laplace, weight and defining-function operators on Fefferman--Graham ambient spaces. The algebra of these operators underlies conformal geometries.
 We extend those results to include fermions by taking an $\frak{osp}(1|2)$ ``Dirac square root'' of these algebras. The theory is a simple, Grassmann, two-matrix model. Its quantum action is a Chern--Simons
 theory whose differential is a first-quantized, quantum mechanical BRST operator. The theory is a basic ingredient for
 building fundamental theories of  physical observables.
 \end{abstract}


\maketitle

\section{Introduction}

Our aim is to construct quantum gravity  models based on two key premises:
\begin{itemize}
\item[(i)] \label{ONE} Physics is the art of pre- and post-dicting the future and past--the first step to construct quantum gravity is to average over all possible causal structures. 
\item[(ii)] \label{TWO} Physics is observation-based--the basic data for a theory of quantum gravity should be an algebra of observables.
\end{itemize}
A first concrete step in this direction was taken in seminal work of Bars and collaborators: 
They wrote down equations  
that predicted the Hamiltonian of  quantum mechanics and constructed an action principle that could be used to quantize these equations~\cite{BarsRey}. Remarkably, they found that the moduli space of their equations was labeled by Fefferman--Graham (FG) metrics. These are $(d+2)$-dimensional ambient metrics that are in correspondence to $d$-dimensional conformal geometries and hence spacetime conformal structures~\cite{FG}.

Of course, it is hard to imagine that  quantum gravity could be based only on conformal geometries {\it alias} Weyl invariant systems. Here a second crucial observation was made by   Bailey, Eastwood and Gover (BEG): The FG construction realizes conformal geometries as the curved analog of  the conical space of ambient lightlike rays. Solutions to Einstein's equations amount  to curved analogs of conical sections~\cite{BEG}.

The BEG description of Einstein's equations amounts to finding a parallel ambient vector field known as a parallel scale tractor while Bars' approach amounts to quantizing 
algebras generalizing the
Laplace, weight and defining function operators on an FG ambient space. [These ${\frak sp}(2)$ ``GJMS algebras''
first arose in a conformal geometry context in a study of invariant Laplacians by
Graham, Jennes, Mason and Sparling~\cite{GJMS}.]
Recently, these two approaches were melded in a study of quantum gravities obtained by coupling scale in the BEG sense  to the Bars' quantized conformal geometries~\cite{Bonezzi}.

The Klein--Gordon operator is fundamental for physics but is 
underpinned by the Dirac operator. 
In this letter, we quantize a certain ``square root'' of the GJMS algebra. This leads to the simplest of this class of quantum gravities, namely an infinite dimensional two-matrix model. The model enjoys a gigantic gauge symmetry, since in some sense it contains infinite towers of interacting higher spins. However, its quantum action is a simple Chern--Simons theory obtained using the geometry of Batalin--Vilkovisky (BV) quantization~\cite{BV} 
discovered by Alexandrov, Kontsevich, Schwarz and Zaboronsky (AKSZ)~\cite{AKSZ}.

\section{The massless Dirac equation} 

The massless Dirac equation
$$
\gamma^\mu \nabla_\mu \psi=0
$$
is Weyl invariant in any dimension~$d$ and curved spacetime~$M$, under local metric and spinor transformations
$$
g_{\mu\nu}\mapsto \Omega^2g_{\mu\nu}\, ,\qquad \psi\mapsto \Omega^{\frac{1-d}{2}}\psi\, .
$$
In other words, the Dirac operator 
is conformally invariant~\footnote{Here we use the mathematical definition of conformal invariance, which refers to covariance under Weyl transformations.}. As observed by Dirac, this equation can be rewritten in a $(d+2)$-dimensional conformal space~\cite{Dirac}, whose curved analog was discovered by Fefferman and Graham by demanding that the ambient, signature $(d,2)$, metric
obeyed
\begin{equation}\label{FG}
g_{MN}=\nabla_M X_N\, ,
\end{equation}
 where here $\nabla$ is the ambient Levi-Civita connection~\cite{FG}.
The closed homothety $X_M$ generates dilations whose eigenvalues are conformal weights, while the zero locus of its square~$X^2$ defines the curved conformal cone. The massless Dirac equation then corresponds to ambient spinors~$\Psi\in{\bf S}M $ subject to
$$
S^+ \Psi = 0 = S^- \Psi\, ,
$$
where $\{\Gamma_M,\Gamma_N\}=2g_{MN}$ 
and \begin{equation}\label{Spm}
S^+:=\slashed{X}\, , \qquad S^-:=\slashed \nabla\, .
\end{equation}
Our goal is not to reformulate the Dirac equation~\footnote{Ambient space tensors are known as tractors~\cite{Gover}. The tractor description of spinors  and supersymmetric systems was given in~\cite{Branson} and~\cite{Shaukat}.}, but rather to explore quantum gravity by probing the space of {\it all} possible Dirac operators. We begin  with a maneuvre reminiscent of string theory's nascent, first-quantized steps.

\section{First Quantized Dirac Equation}

To describe the Dirac equation ambiently in first quantization, one first notes that  the operators $S^{\pm}$ generate a first class, ${\mathfrak osp}(1|2):=\{S^{\pm},Q^{\pm\pm},Q^{+-}\}$ constraint algebra~\cite{Holland} where 
\begin{eqnarray}
&Q^{++}:=(S^+)^2=X^2\, ,\quad
\!\!Q^{--}:=(S^-)^2=\Delta-\frac14{R}\, ,&
\nonumber\\[1mm]
&Q^{+-}:=\{S^+,S^-\}=2\nabla_X+d+2\,  . &\label{Qs}
\end{eqnarray}
We shall call this a {\it conformal Dirac algebra} (CDA).

The worldline particle model~\footnote{The bosonic analog of this model was first proven to be equivalent to the relativistic paricle by Marnelius~\cite{Marnelius} and
then employed as the basis of a ``two-times physics'' program in~\cite{Bars}.} $$S=\int\big(P_M\dot X^M-[\lambda_i S^i+\lambda_{ij} Q^{ij}]\big)$$ imposes the CDA constraints in Dirac quantization. By making differing gauge choices, it describes various models. These include  
relativistic and constant curvature spinning particles, the hydrogen atom with spin and other $SO(d,2)$ invariant conformal  models~\cite{Bars:1998gv}.
Its worldline BRST operator can be treated using the detour methods of~\cite{Cherney}. This yields an equivalent, reduced BRST operator 
$$
Q_{\rm BRST}={\bf d}+q \big(2\nabla_X+d+2\big)+z \slashed X +
\slashed\nabla \partial_p \, ,
$$
which acts on ambient spinor, worldline $(z,p,q)$-ghost-polynomial wavefunctions living in the subspace ${\rm coker}(z^2)\cap{\rm ker}(\partial_p^2)=:{\mathcal H}_{\rm BRST}$. The reduced Lie algebra differential 
\begin{equation}\label{d}
{\bf d}:=q(1-z\partial_z-p\partial_p)-z\partial_p\partial_q
\end{equation}
is separately nilpotent acting on ${\mathcal H}_{\rm BRST}$.
The BRST cohomology  is that of the ghost-number-graded complex 
$$
0\rightarrow {\bf S}M
\stackrel{\!\scalebox{.5}{$\begin{pmatrix}S^+\\ S^-\end{pmatrix}$}}{\longrightarrow}
{\bf S}M^{\otimes 2}
\stackrel{\!\scalebox{.45}{$\begin{pmatrix}S^- S^+ \!-2\!&- S^+S^+\\[2mm]S^-S^-&\!-S^+S^-\!-2\end{pmatrix}$}}{
-\!\!\!-\!\!\!-\!\!\!-\!\!\!-\!\!\!-\!\!\!-\!\!\!\longrightarrow}
{\bf S}M^{\otimes 2}
\stackrel{\!\scalebox{.5}{$\begin{pmatrix}-S^-&S^+\end{pmatrix}$}}{-\!\!\!-\!\!\!\longrightarrow}
{\bf S}M\rightarrow 0\, .
$$
Massless spinors form the ghost number zero cohomology.

\section{Second quantization}

In second quantization the operators $S^\pm$ are  off-shell and obey  equations of motion. These are  exactly the integrability conditions required for  the above sequence of maps to be a complex, namely
$$
[S^-, S^+S^+] -2S^+=0=[S^-S^-,S^+] -2S^-\, .
$$
These equations follow from the action principle
\begin{equation}\label{Corradini}
S_{\rm cl}={\rm tr}\Big(S^+S^-\,\! +\, \frac12\,  S^+ S^+ S^-S^-\Big)\, .
\end{equation}
This is a simple, Grassmann, two-matrix model except that the trace is over spinor bundle~${\bf S}M$ operators. 

The above integrability conditions ensure that the composite operators $(S^\pm,S^\pm S^\pm,\{S^+,S^-\})$ obey an ${\frak osp}(1|2)$ algebra. Indeed, one can ``integrate in'' new ``fields'' $Q^{ij}$ and finds an equivalent cubic action principle
$$
S_{\rm cl}={\rm tr}\Big(\, \frac12\, S^iS_i 
+\frac12 \, Q^{ij}Q_{ij}-\frac13 \, S^i Q_{ij}S^j
\Big)\, .
$$
This is our classical action principle; non-trivial solutions
include the Dirac operator multiplet given in
Equations~\eqref{Spm} and~\eqref{Qs}.  These solutions rely on the FG metric condition~\eqref{FG}. This shows that the moduli space is parameterized, in part, by conformal geometries and hence causal structures. Moreover our space of observables is the algebra 
of operators acting on ambient spinors.

\section{Quantum action}

The classical action~\eqref{Corradini}
enjoys a huge gauge invariance
$$
\delta S^\pm=[S^\pm,\varepsilon]\, .
$$
For example, expanding the operator valued parameter $$\varepsilon=\epsilon+\xi^M\nabla_M + \zeta^{MN}\nabla_M\nabla_N + \cdots\, ,$$ the ambient fields $(\epsilon, \xi^M)$ parameterize Maxwell and ambient diffeomorphism invariances while $\zeta^{MN}$ is the parameter for the first of an infinite tower of higher spin gauge symmetries~\cite{BarsRey}.

To quantize the theory we must construct its quantum action. We start by second quantizing the 
worldline BRST Hilbert space ${\mathcal H}_{\rm BRST}\ni {\mathcal A}$, which means that a wavefunction~${\mathcal A}$ becomes a field and thus, in this context, an operator on ${\bf S}M$.
In modern BV language,~${\mathcal A}$ is a coordinate for an infinite dimensional $Q$-manifold~\cite{SchwarzQ} whose differential is given by ${\bf d}$ of Equation~\eqref{d}. 

Expanding the polynomial ${\mathcal A}$ in powers of $(z,p,q)$ determines the BV field content:
\begin{equation*}
\begin{split}
{\mathcal A}&:=S^+\!\!+z\lambda^*\!+pC^*\!+zp S^-\!
+
q\big(S_+^*\!+zC\!+p\lambda+zpS_-^*\big) .
\end{split}
\end{equation*}
The quantum action is a Chern--Simons theory~\footnote{It is natural to conjecture that the model can equivalently be formulated, at the cost of an infinite tower of auxiliary fields, in terms of unrestricted polynomials ${\mathcal A}(z,p)$ where the ``fields'' $Q^{ij}$ are off-shell.}
\begin{equation}\label{Squ}
S_{\rm qu}={\rm tr}\int \Big({\mathcal A}{\bf d}{\mathcal A}+\tfrac23 \tfrac{\partial}{\partial p} 
{\mathcal A}^3\Big)\, .
\end{equation}
Here, the integral denotes the ghost measure given by projection onto monomials proportional to $zpq$.
The partial~$p$-derivative is thus {\it not} a total derivative, but rather defines a cyclic triple product.
By construction, the above action enjoys an enhanced gauge invariance
$$
\delta {\mathcal A}={\bf d}_{\mathcal A} {\mathcal E}:={\bf d} {\mathcal E}+ \tfrac\partial{\partial p} [{\mathcal A},{\mathcal E}]
\, ,
$$
which is fixed by path integrating over any (odd) Lagrangian submanifold of the underlying $Q$-manifold~\cite{Schwarzfix}. In the above formula, the operator $\partial_p$ is required both on grounds of ghost number and requiring that the commutator of field and parameters lives in ${\rm ker}(\partial_p^2)$.

The quantum action~\eqref{Squ} is a sum of a classical action
$$
S_{\rm cl}={\rm tr} \Big(
S^+S^--\frac12\, \lambda^2-\lambda\, \{S^+,S^-\}\Big)\, ,
$$
plus the standard minimal BV terms built from antifields multiplying the corresponding  BRST variations of fields:
$$
s S^{\pm }=\{C,S^{\pm}\}\, ,\quad
s \lambda = [C,\lambda]\, ,\quad
s C= C^2\, .
$$
Notice that there is an additional ``gauge field'' $\lambda$. This is an auxiliary 
field corresponding to the operator $Q^{+-}$; when $\lambda$ is integrated out classically we recover the action~\eqref{Corradini}.

At the quantum level, the model's path integral sums over all possible Dirac operators whose on-shell moduli space corresponds to conformal geometries and so, as promised, the model is a weighted average over causal structures.

\section{Conclusions}

Our model is closely related to other leading candidates for quantum gravity theories, in particular string field theory~\cite{SFT} and Vasiliev's higher spin theory~\cite{Vasiliev}.
String theory is finite, anomaly- and tachyon-free. However it is far from clear that 
this holds for either the present model or the Vasiliev theory (see however~\cite{Boulanger}), although at least the field content of the latter model is already well-understood (see for example~\cite{Review}). Addressing these gaps is an
obvious future research direction.

The presence of a graviton is at least easier to understand using the parallel scale tractor of~\cite{BEG}; the key is to couple conformal geometry to scale. For example, the action principle $S=\int \lambda^{ij}
Q_{ij} \psi$ is known to be gauge equivalent to the Einstein--Hilbert action~\cite{BarsBRST,BonezziO}. In~\cite{Bonezzi}, coupling to scale was achieved by supersymmetrizing the algebra of observables. The Hilbert space trace became an ${\mathcal N}=2$ supertrace and in turn,  the model there was found to have a graviton in its spectrum. Coupling our model to scale should similarly yield a propagating graviton.

Another key question is the computation of physical correlators. 
Here the advantage of the BV and AKSZ methods comes to the fore; observables can be viewed as  the homology of Lagrangian submanifolds of the $Q$-manifold~\cite{Schwarzfix,Sezgin}.
In a similar vein, perturbation theory is also (in principle) simple, because the BV propagator 
is just $\frac{\bm \delta}{\bm \Delta}$ where  ${\bm \delta}$ and ${\bm \Delta}$ are the (worldline) antiBRST differential and BRST Laplacian, respectively~\cite{Schwarzfix} (see also~\cite{Bonezzi}).

Despite the unanswered questions listed above, the model has some compelling features. Albeit infinite dimensional, it is a simple matrix model, and thus conceivably finite--the theory can be regulated using well-studied (see~\cite{Marino}) matrix models. Moreover, unlike string and Vasiliev theories, the
cubic product describing the joining and splitting of Hilbert spaces is very simple, which may augur well for model building. In particular, the extension to form fields based on an ${\mathfrak osp}(2|2)$ algebra is immediate.

\begin{acknowledgments}
We thank Itzhak Bars, Rod Gover, Robin Graham, Albert Schwarz and Per Sundell for discussions.
R.B. and A.W. thank the Universidad Andres Bello for hospitality.
E.L. acknowledges partial support from SNF Grant No. 200020-149150/1.
A.W. and O.C. were supported in part by the 
UCMEXUS-CONACYT grant CN-12-564. 
A.W. was supported in part by a Simons Foundation Collaboration Grant for Mathematicians.
\end{acknowledgments}


\end{document}